\documentclass[apj,twocolumn,letter,twocolappendix]{aastex6}

\usepackage[encapsulated]{CJK}
\usepackage{ucs}
\usepackage[utf8x]{inputenc}
\newcommand{\cntext}[1]{\begin{CJK}{UTF8}{gbsn}#1\end{CJK}}

\shorttitle{Hot gas in the WR nebula NGC\,3199}
\shortauthors{Toal\'{a} et al.}

\usepackage{rotating}
\usepackage{color}

\begin{document}

\title{Hot gas in the Wolf-Rayet nebula NGC\,3199}

\author{J.A.\,Toal\'{a}\,\cntext{(杜宇君)}$^{1,2}$}
\author{A.P.\,Marston$^{3}$}
\author{M.A.\,Guerrero$^{4}$}
\author{Y.-H.\,Chu\,\cntext{(朱有花)}$^{1}$}
\author{R.A.\,Gruendl$^{5}$}

\affil{$^{1}$Institute of Astronomy and Astrophysics, Academia Sinica\,(ASIAA),
  Taipei 10617, Taiwan\\
$^{2}$Instituto de Radioastronom\'{i}a y Astrof\'{i}sica, UNAM Campus Morelia, Apartado postal 3-72, Morelia 58090, Michoac\'{a}n, Mexico\\  
$^{3}$European Space Agency/STScI, 3700 San Martin Drive, Baltimore, MD 21218, USA\\
$^{4}$Instituto de Astrof\'\i sica de Andaluc\'{i}a, IAA-CSIC,
  Glorieta de la Astronom\'\i a s/n, Granada 18008, Spain\\
$^{5}$Department of Astronomy, University of Illinois,
  1002 West Green Street, Urbana, IL 61801, USA}

\begin{abstract}
  The Wolf-Rayet (WR) nebula NGC\,3199 has been suggested to be a bow
  shock around its central star WR\,18, presumably a runaway star,
  because optical images of the nebula show a dominating arc of
  emission south-west of the star. We present the {\it XMM-Newton}
  detection of extended X-ray emission from NGC\,3199, unveiling the
  powerful effect of the fast wind from WR\,18. The X-ray emission is
  brighter in the region south-east of the star and analysis of the
  spectral properties of the X-ray emission reveals abundance
  variations: i) regions close to the optical arc present
  nitrogen-rich gas enhanced by the stellar wind from WR\,18 and ii)
  gas at the eastern region exhibits abundances close to those
  reported for nebular abundances derived from optical studies,
  signature of an efficient mixing of the nebular material with the
  stellar wind. The dominant plasma temperature and electron density
  are estimated to be $T\approx1.2\times$10$^{6}$~K and
  $n_\mathrm{e}$=0.3~cm$^{-3}$ with an X-ray luminosity in the
  0.3--3.0~keV energy range of
  $L_\mathrm{X}$=2.6$\times$10$^{34}$~erg~s$^{-1}$. Combined with
  information derived from {\it Herschel} and the recent {\it Gaia}
  first data release, we conclude that WR\,18 is not a runaway star
  and the formation, chemical variations, and shape of NGC\,3199
  depend on the initial configuration of the interstellar medium.
\end{abstract}


\maketitle

\section{INTRODUCTION}
\label{sec:intro}

Throughout their lives, before exploding as supernovae, very massive
stars ($M_\mathrm{i} \gtrsim$~30 ~M$_{\odot}$) modify the structure
and enrich the interstellar medium (ISM) by a combination of different
factors: stellar winds, strong ionizing photon fluxes, and proper
motions. They represent the main source of feedback that governs the
physical structures of the ISM.

After evolving off the main sequence stage, very massive stars enter
the red supergiant or luminous blue variable phase
exhibiting dense, slow ($\sim$10--100~km~s$^{-1}$), and dust-rich
winds that expand into the ISM. This slow wind expels more than half
of the initial mass of the star, exposing its hot core and becoming a
Wolf-Rayet (WR) star. WR stars present strong winds
\citep[$v_{\infty}\approx 1500$~km~s$^{-1}$,
  $\dot{M}\approx$10$^{-5}$~M$_{\odot}$~yr$^{-1}$; e.g.,
][]{Hamann2006} that sweep up and compress the previously ejected
RSG/LBV material into a shell, whilst the newly developed UV flux
ionizes the material, forming the so-called ring nebulae or WR
nebulae.

Bubble models suggest that this wind-wind interaction produces an
adiabatically-shocked region of gas with temperatures as high as
$T$=10$^{7}$--10$^{8}$~K and electron densities of
$n_\mathrm{e}\approx$10$^{-2}$~cm$^{-3}$ that fills the nebular shell
interior \citep[][]{Dyson1997}, known as {\it hot bubble}\footnote{The
  post-shock temperature in a hot bubble can be expressed as
  $k_\mathrm{B} T = 3 \mu m_\mathrm{H} v_{\infty}^{2} / 16$, where
  $\mu m_\mathrm{H}$ and $k_\mathrm{B}$ are the mean mass per particle
  and the Boltzmann's constant, respectively.}. These hot bubbles have
only been detected by X-ray observatories in the WR nebulae around
WR\,6 (S\,308), WR\,7 (NGC\,2359), and WR\,136 (NGC\,6888). The
best-quality X-ray observations towards these nebulae have been
obtained with {\it XMM-Newton} and model fits to their X-ray spectra
suggest plasma temperatures of $T_\mathrm{X}$=[1--2]$\times10^{6}$~K,
electron densities $n_\mathrm{e}\lesssim1$~cm$^{-3}$, and X-ray
luminosities $L_\mathrm{X}=$10$^{33}$--10$^{34}$~erg~s$^{-1}$
\citep[][]{Chu2003,Toala2012,Toala2015,Toala2016}. The low
temperatures indicated by the soft diffuse X-ray emission are the
result of mixing between the hot bubble and the warm ($T$=10$^{4}$~K)
nebular material. Properties of this mixing region stongly depend on
the formation of hydrodynamical instabilities and can be augmented if
thermal conduction is taken into account
\citep[e.g.,][]{Toala2011,Dw2013}. Consequently, the X-ray-emitting
gas exhibits abundances close to those of the nebular material.

\begin{figure*}
\begin{center}
\includegraphics[angle=0,width=0.9\linewidth]{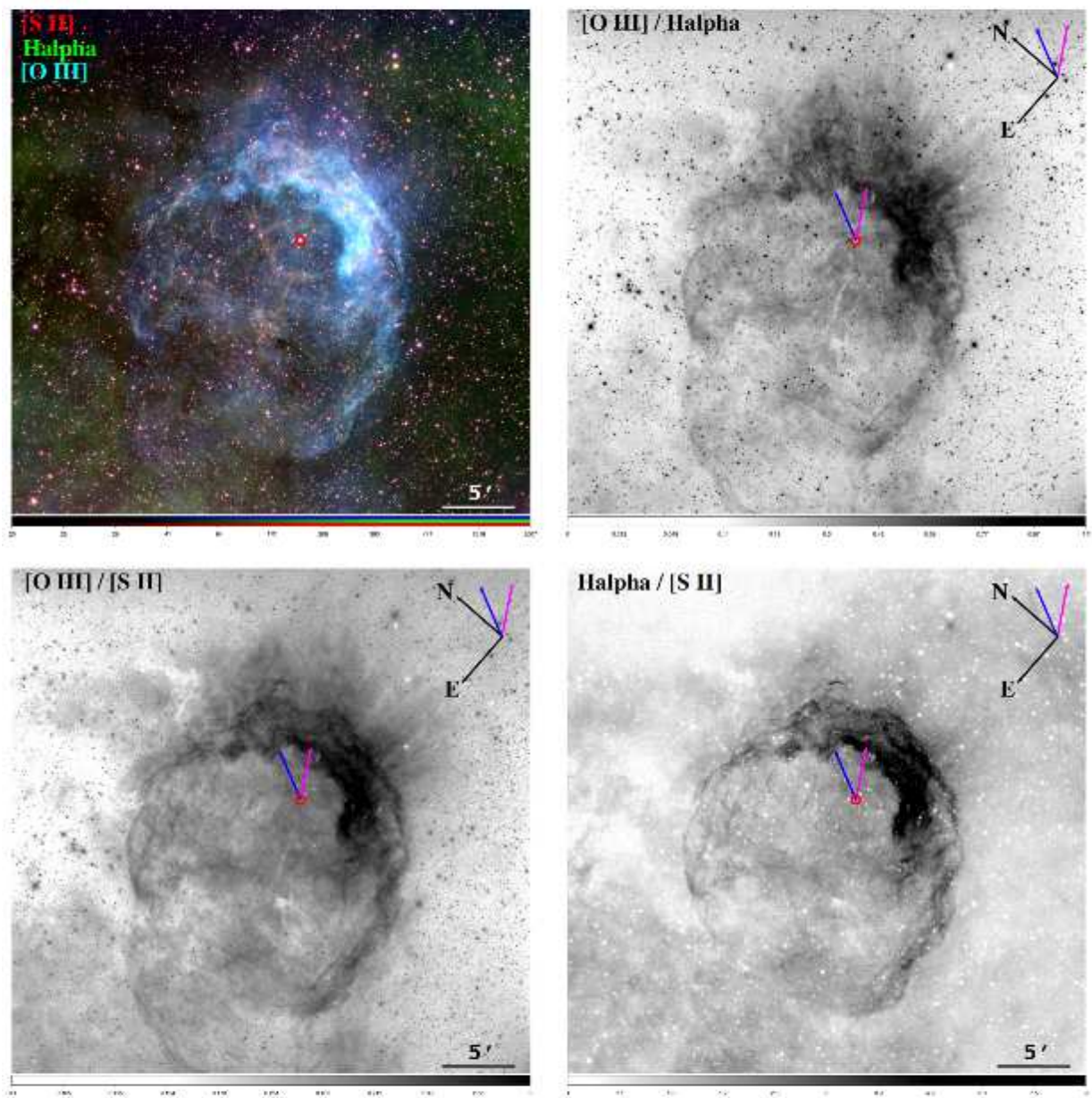}
\label{fig:NGC3199_don}
\caption{Top left panel: Color-composite nebular image of
  NGC\,3199. The colors red, green, and blue correspond to [S\,{\sc
      ii}], H$\alpha$, and [O\,{\sc iii}] line emission,
  respectively. Other panels show the [O\,{\sc iii}]/H$\alpha$
  (top-right), [O\,{\sc iii}]/[S\,{\sc ii}] (bottom-left), and
  H$\alpha$/[S\,{\sc ii}] (bottom-right) ratio maps. The position of
  WR\,18 is shown with a red circle in all panels and the orientation
  of the figures is shown on each panel. The blue and magenta arrows
  show the direction of the proper motions reported by {\it Hipparcos}
  and {\it Gaia} observations, respectively (see text). The
  narrow-band images are courtesy of Don Goldman.}
\end{center}
\end{figure*}

The deepest X-ray observation of a WR nebula is that presented by
\citet{Toala2016} of NGC\,6888, the most studied object of this
class. These authors reported, for the first time in the X-ray regime,
nitrogen and temperature variations within the WR nebula leading them
to the conclusion that the mixing of material is not equally efficient
in all directions. Apparently, the mixing has been less efficient
towards the caps of NGC\,6888, which presents higher temperature and
nitrogen abundance, while in the central regions the X-ray-emitting
material has similar abundances as those reported for the nebular
material \citep[see][and references therein]{ReyesPerez2015}.

On the other hand, there are two other WR nebulae that have not been
detected in X-rays. These are RCW\,58 around WR\,40 and that around
WR\,16 \citep{Gosset2005,Toala2013}. These nebulae harbor WN8h stars
with relatively slow stellar winds
($v_{\infty}\approx650$~km~s$^{-1}$), while those that have been
detected in X-rays (S\,308, NGC\,2359, and NGC\,6888) have WN4-6 stars
with faster winds
\citep[$v_{\infty}\approx1700$~km~s$^{-1}$;][]{Hamann2006}. Even
though this wind velocity seems to be the only characteristic in
common for displaying diffuse X-ray emission, it can not be taken with
certainty as the current number of studied WR nebulae in the X-ray
regime is small and the global properties of the X-ray-emitting gas
may depend on other stellar and nebular properties, including the
stellar evolution, formation mechanism, ISM structures around the
nebulae \citep[see discussion by][]{Toala2016}.

In this paper, we present {\it XMM-Newton} observations towards
NGC\,3199 (see Fig.~\ref{fig:NGC3199_don}) around the WN4 star WR\,18
\citep[$T_\mathrm{eff}=112.2$~kK;][]{Hamann2006}. Archival {\it
  ROSAT} observations (Obs.\,ID.\,RP900318N00) hinted at the presence
of diffuse X-ray emission, but the X-ray point sources projected
within the nebula, including that of WR\,18
\citep[e.g.,][]{Skinner2010}, prohibited an unambiguos
analysis. Unlike other WR nebulae detected in X-rays, this nebula has
been the center of discussions as some authors suggest that the
abundances are close to galactic H\,{\sc ii} regions \citep[see,
  e.g.,][]{Whitehead1988,Stock2011}, while others suggest that some
regions within the nebula exhibit significant chemical enrichment
\citep[e.g.,][]{Esteban1992,Marston2001}.

Figure~1 shows in great detail the optical morphology and extension of
NGC\,3199. The nebula has an elongated shape with angular size of
$\sim$20\arcmin$\times$25\arcmin\, as mapped by the [O\,{\sc iii}]
narrow-band emission but with an obvious enhanced emission towards the
west, the bright arc \citep[see also][]{Whitehead1988}. This arc was
the defining factor for \citet{Dyson1989} to suggest that NGC\,3199 is
composed of swept up ISM material in a bow shock around WR\,18, which
is not located at the geometrical center of the nebula. In fact, {\it
  Hipparcos} reported a proper motion for WR\,18 of ($\mu_{\alpha}$,
$\mu_{\delta}$)=($-$2.30$\pm$1.75, 4.78$\pm$1.35~mas~yr$^{-1}$)
\citep{Perryman1997}, which at a distance of 2.2~kpc
\citep[see][]{vdHucht2001} corresponds to a projected velocity of
$v_{\star} \approx55\pm$20~km~s$^{-1}$ along the north-west direction.

The present paper is organized as follows. The description of our {\it
  XMM-Newton} observations and the analysis are presented in
Section~2. Our results are presented and discussed in Sections 3 and
4, respectively. We finally summarize our findings in Section~5.

\section{OBSERVATIONS AND DATA PREPARATION}
\begin{figure*}
\begin{center}
\includegraphics[angle=0,width=0.49\linewidth]{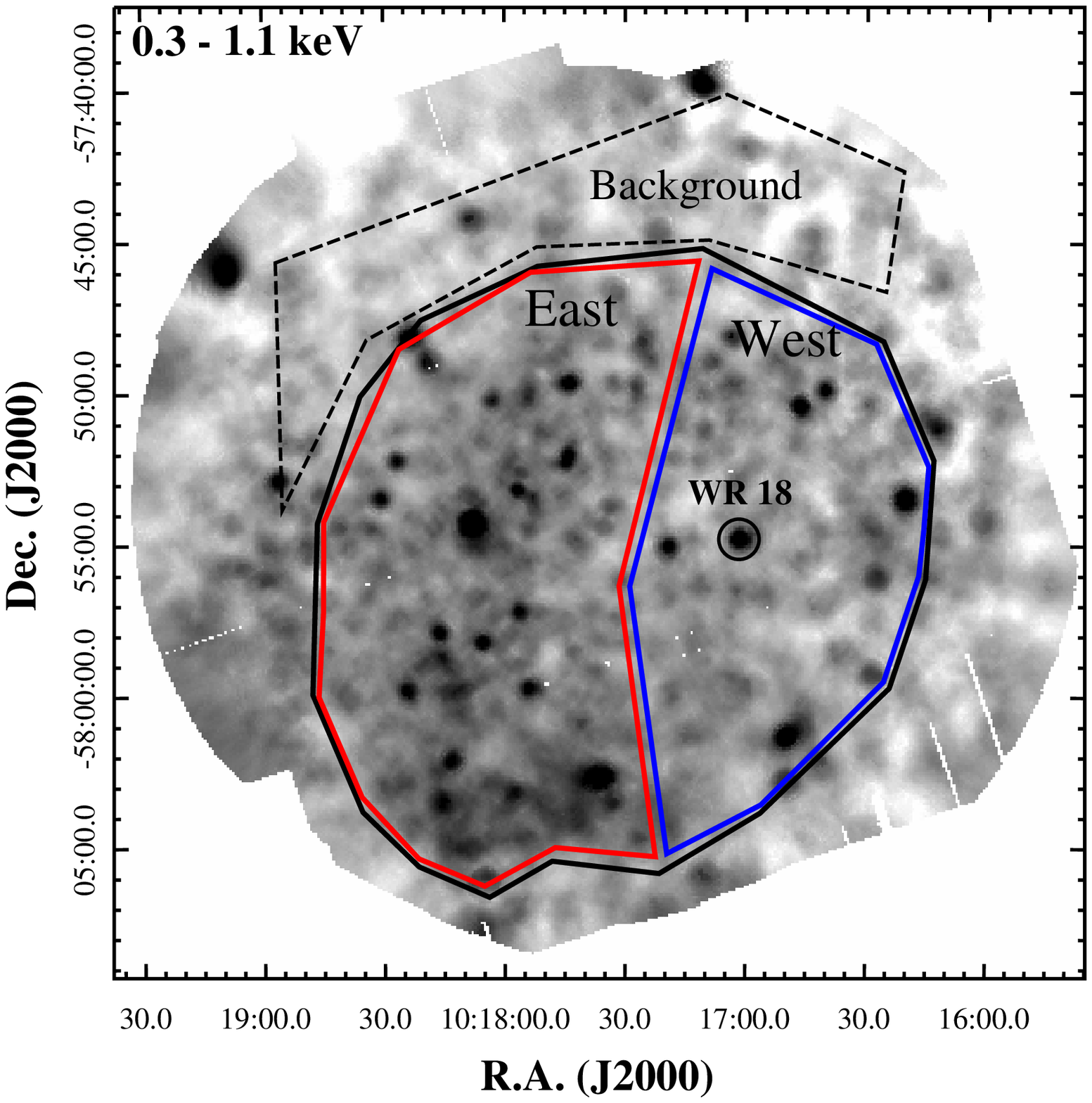}~
\includegraphics[angle=0,width=0.49\linewidth]{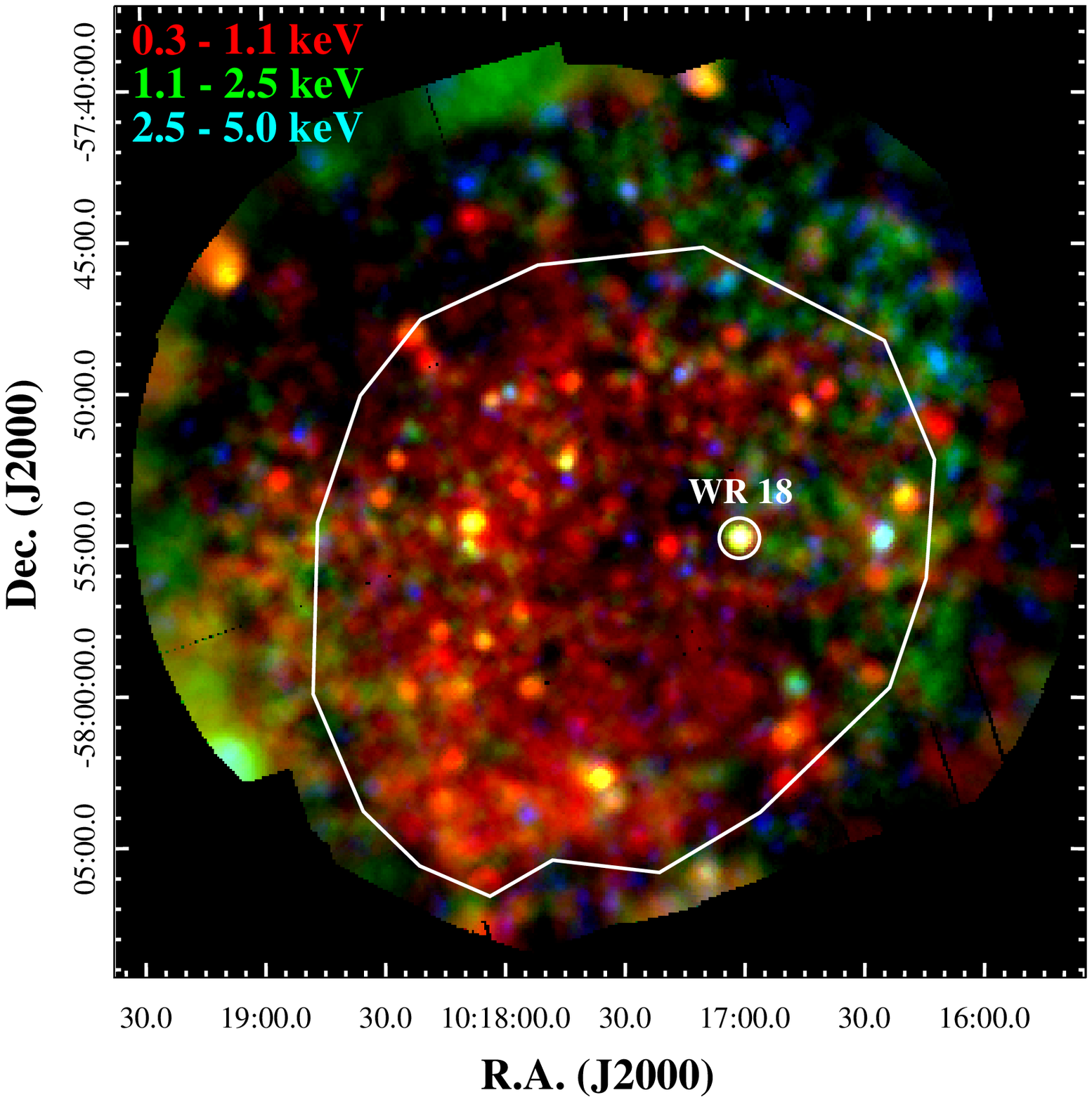}
\label{fig:ESAS}
\caption{{\it XMM-Newton} EPIC (MOS1+MOS2+pn) exposure-corrected
  images of the X-ray emission from NGC\,3199. Left: Soft band
  (0.3--1.1~keV). Right: Color-composite X-ray image. The colors red,
  green, and blue correspond to the soft, medium, and hard X-ray
  bands, respectively. The central star, WR\,18, is shown with a
  circular aperture in both panels. The regions used for source
  spectral extraction are shown by solid lines whilst the background
  region by dashed line. No point sources have been excised from these
  images.}
\end{center}
\end{figure*}

The WR nebula NGC\,3199 around WR\,18 was observed by the European
Science Agency (ESA) X-ray telescope {\it XMM-Newton} on 2014 December
1 in revolution 2743 (Observation ID: 0744460101; PI:
J.A.\,Toal\'{a}). The European Photon Imaging Camera (EPIC) was used
in the Extended Full Frame Mode with the Medium Optical Filter. The
total time of the observation was 56.8~ks with exposure times of 51.4,
53.4, and 53.2~ks for the EPIC-pn, EPIC-MOS1, and EPIC-MOS2,
respectively.

We processed the EPIC observations using the {\it XMM-Newton} Science
Analysis Software (SAS) Version 15.0 with the corresponding
Calibration Access Layer obtained on 2016 February 26. The event files
have been produced from the Observation Data Files by using the tasks
{\it epproc} and {\it empproc} included in SAS. We identified periods
of high-background level by creating light curves in the 10--12~keV
energy range with a binning of 100~s for each of the EPIC cameras. We
rejected times with count rates higher than 0.3~counts~s$^{-1}$ for
the EPIC-pn camera and 0.2~counts~s$^{-1}$ for the MOS cameras. The
resulting useful exposure times for the pn, MOS1, and MOS2 cameras are
36.1, 48.9, and 49.1~ks, respectively.

\subsection{Spatial distribution of X-rays in NGC\,3199}

To study the spatial distribution of the X-ray-emitting gas in
NGC\,3199, we have used the {\it XMM-Newton} Extended Source Analysis
Software package (XMM-ESAS) included in the current SAS version. We
have followed the ESAS cookbook for the analysis of extended sources
version
5.9\footnote{http://heasarc.gsfc.nasa.gov/docs/xmm/esas/cookbook/xmm-esas.html}
to create maps of the extended X-ray emission in NGC\,3199 and to
identify potential contaminant point sources.

We have created EPIC images in the energy bands 0.3--1.1, 1.1--2.5,
and 2.5--5.0~keV that we label as soft, medium, and hard X-ray bands,
respectively. Individual pn, MOS1, and MOS2 images were extracted,
merged together, and corrected for exposure maps. Figure~2 shows the
resultant background-subtracted EPIC image of the soft band and a
color-composite picture of the three X-ray bands. Each X-ray image has
been adaptively smoothed using the XMM-ESAS task {\it adapt} requiring
80, 80, and 30~counts for the smoothing kernel for the soft, medium,
and hard band, respectively.

\begin{figure*}[ht]
\begin{center}
  \includegraphics[angle=-90,width=1\linewidth]{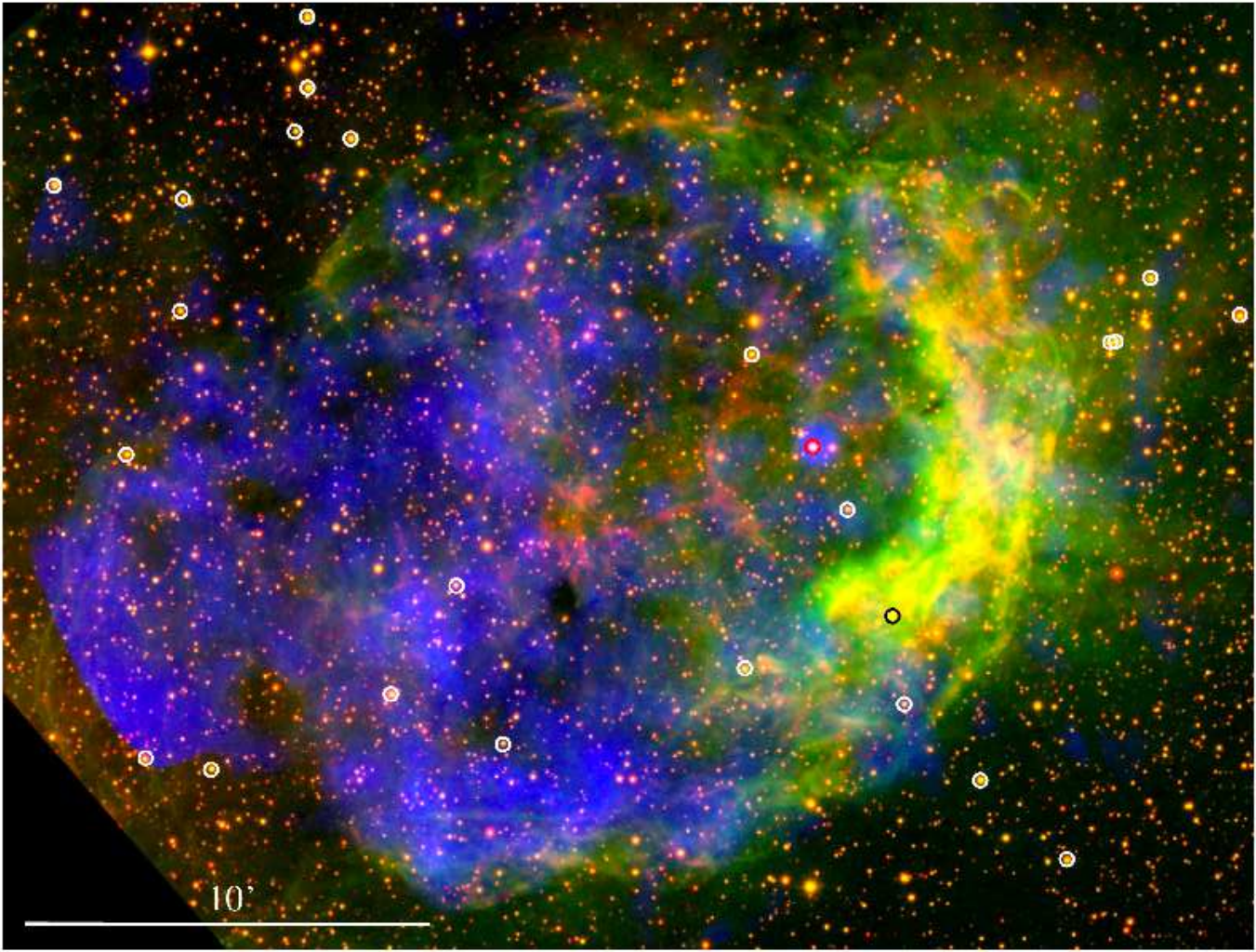}
\label{fig:ESAS_soft}
\caption{Color-composite image of NGC\,3199 around WR\,18. Red, green,
  and blue correspond to [S\,{\sc ii}], [O\,{\sc iii}], and the soft
  X-ray band (0.3--1.1~keV), respectively. All point-like sources have
  been cut out from the X-ray image except WR\,18 (see text for
  details). The position of WR\,18 is shown with a red circle whilst
  other circles mark the positions of the nerby Tycho-Gaia Astrometric
  Solution stars in the field of view. The black circle shows the
  position of the BOV star CD$-$57$^{\circ}$3120 (see
  Discussion). North is up, east to the left.}
\end{center}
\end{figure*}

Figure~2 shows the presence of diffuse X-ray emission within the WR
nebula NGC\,3199, as well as a large number of point-like sources
projected on the nebula. The central star, WR\,18 is detected
confirming the previous {\it Chandra} detection
\citet[][]{Skinner2010}. It seems that the most intense X-ray emission
region is located toward the east of the nebula, in particular, the
south-east region. Furthermore, the X-ray colors unveil spectral
differences within the diffuse X-ray emission: the eastern regions
emit predominantly soft emission (0.3--1.1~keV) while the western
region emits significantly in the medium X-ray band (1.1--2.5~keV). To
further illustrate this, we have used the CIAO {\it dmfilth} routine
\citep[version 4.8,][]{Fruscione2006} to cut out all detected
point-like sources and create a direct comparison between the nebular
emission and the distribution of the X-ray-emitting gas. The
identification of the point sources has been performed following the
EPIC source finding thread\footnote{See
  http://www.cosmos.esa.int/web/xmm-newton/sas-thread-src-find-stepbystep}. This
allowed us to perform a search for point sources in different energy
bands (0.3--0.5~keV, 0.5--1.0~keV, 1.0--2.0~keV, 2.0--4.5~keV, and
4.5--12.0~keV) for the three EPIC cameras.

Figure~3 shows the comparison of the nebular [S\,{\sc ii}] and
[O\,{\sc iii}] narrow-band images (in red and green, respectively)
presented in Fig.~1 with the resultant soft X-ray image (blue). It can
be seen that the diffuse X-ray emission in NGC\,3199 is delimited by
the [O\,{\sc iii}] as in other WR nebulae, filling the cavities
observed in narrow-band nebular line images. Moreover, the extended
X-ray emission is considerably brighter toward the east.

\subsection{Spectral extraction}

\begin{figure*}
\begin{center}
\includegraphics[angle=0,width=0.42\linewidth]{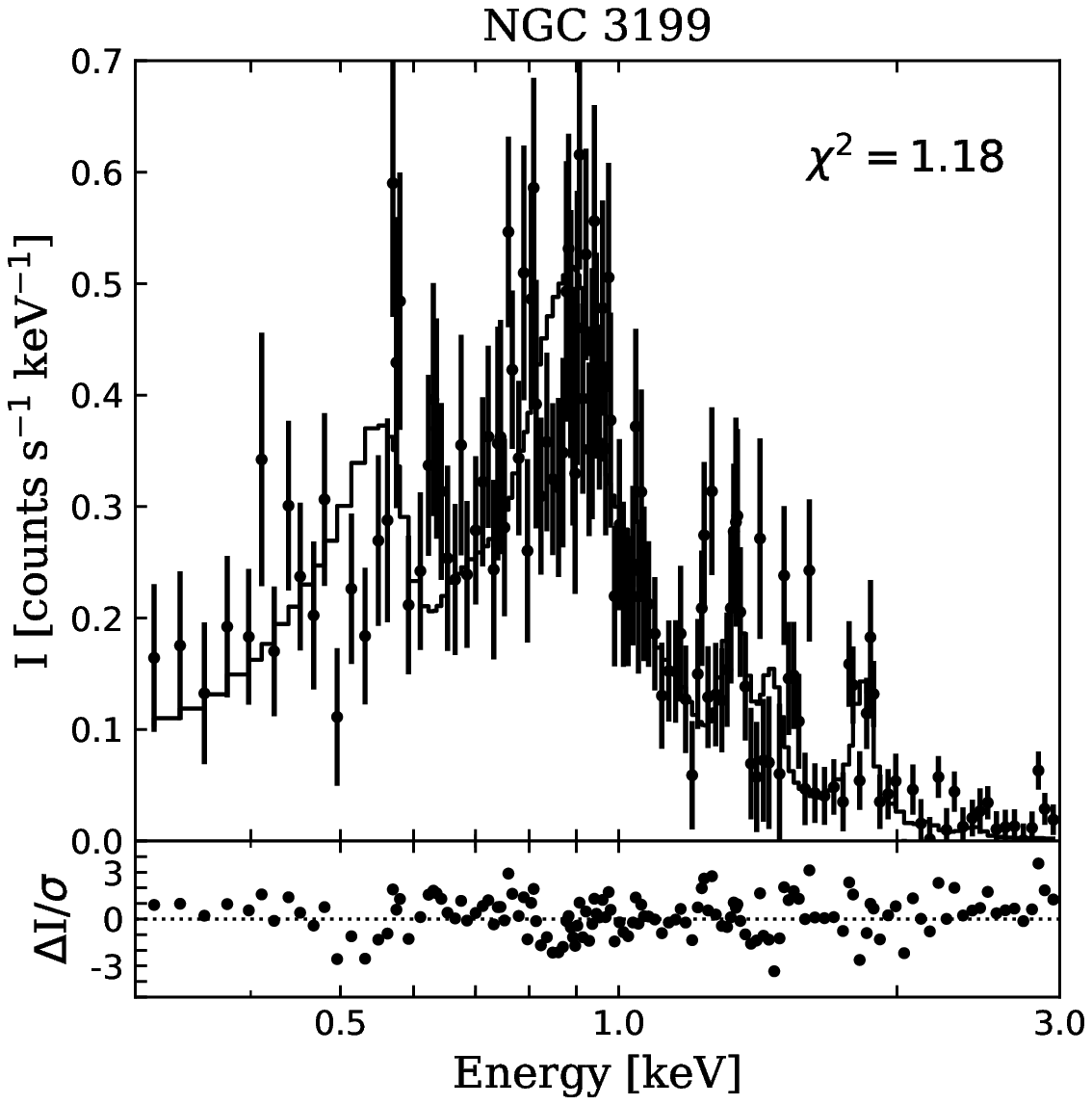}~
\includegraphics[angle=0,width=0.42\linewidth]{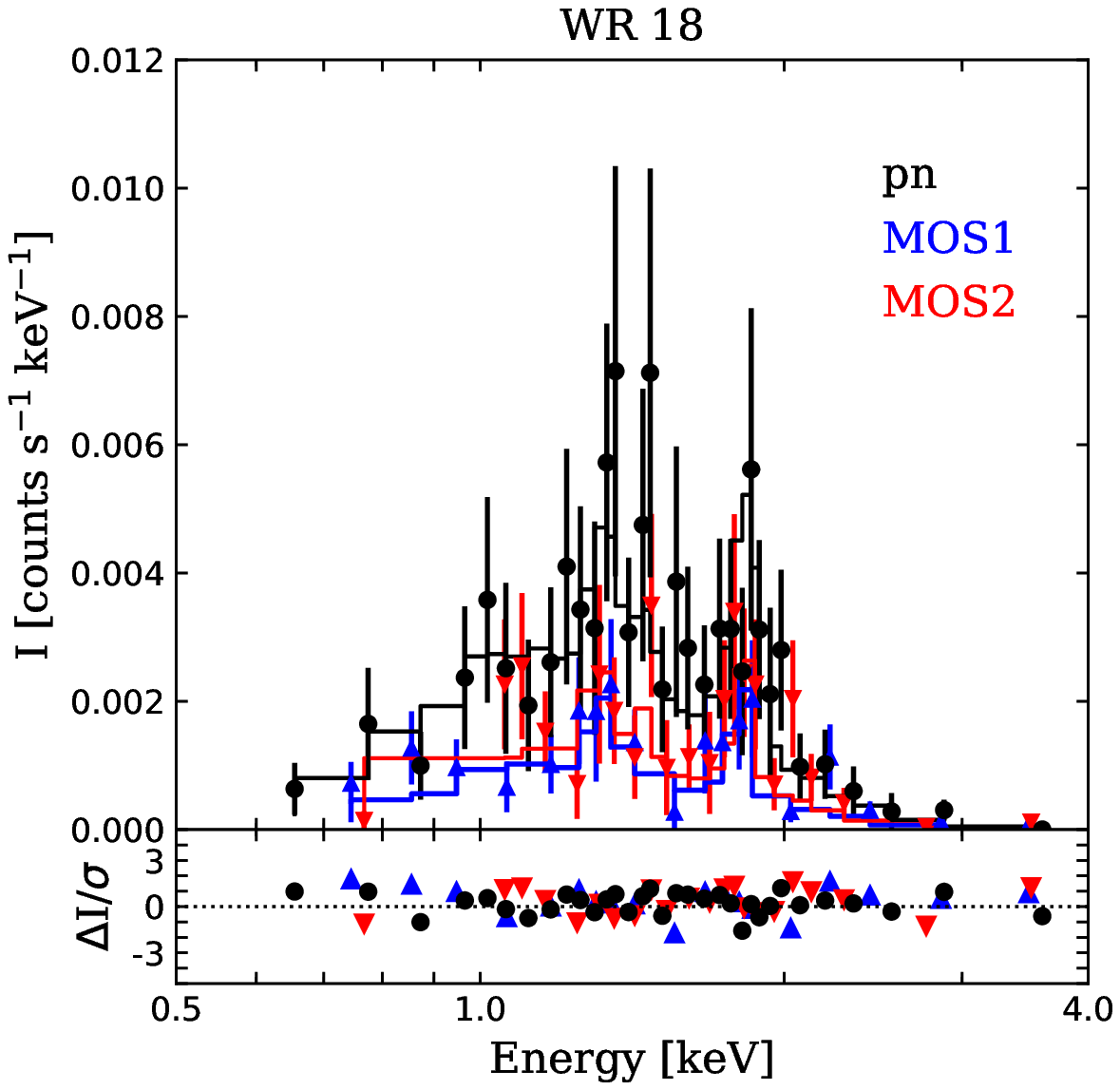}\\
\includegraphics[angle=0,width=0.42\linewidth]{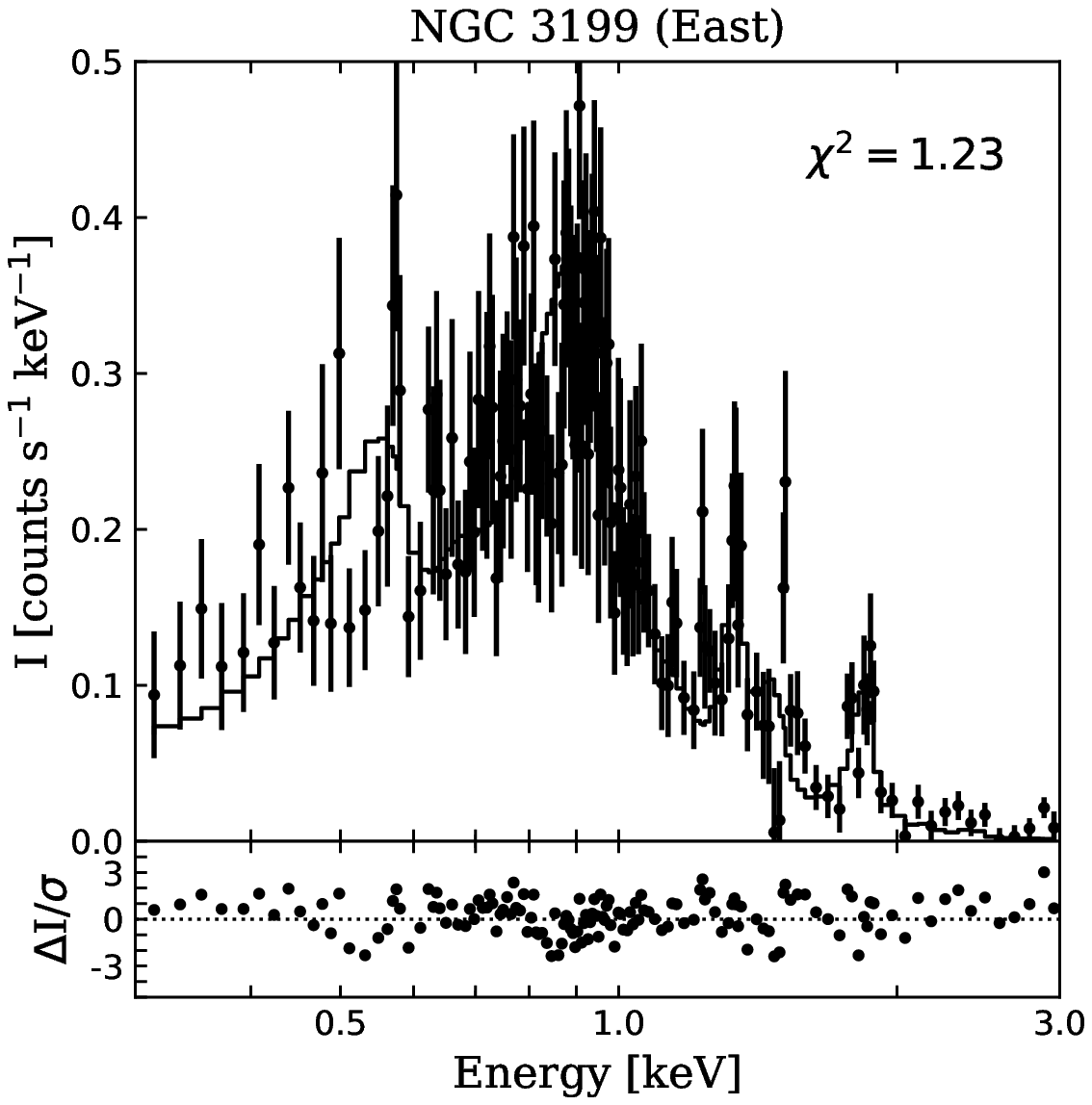}~
\includegraphics[angle=0,width=0.42\linewidth]{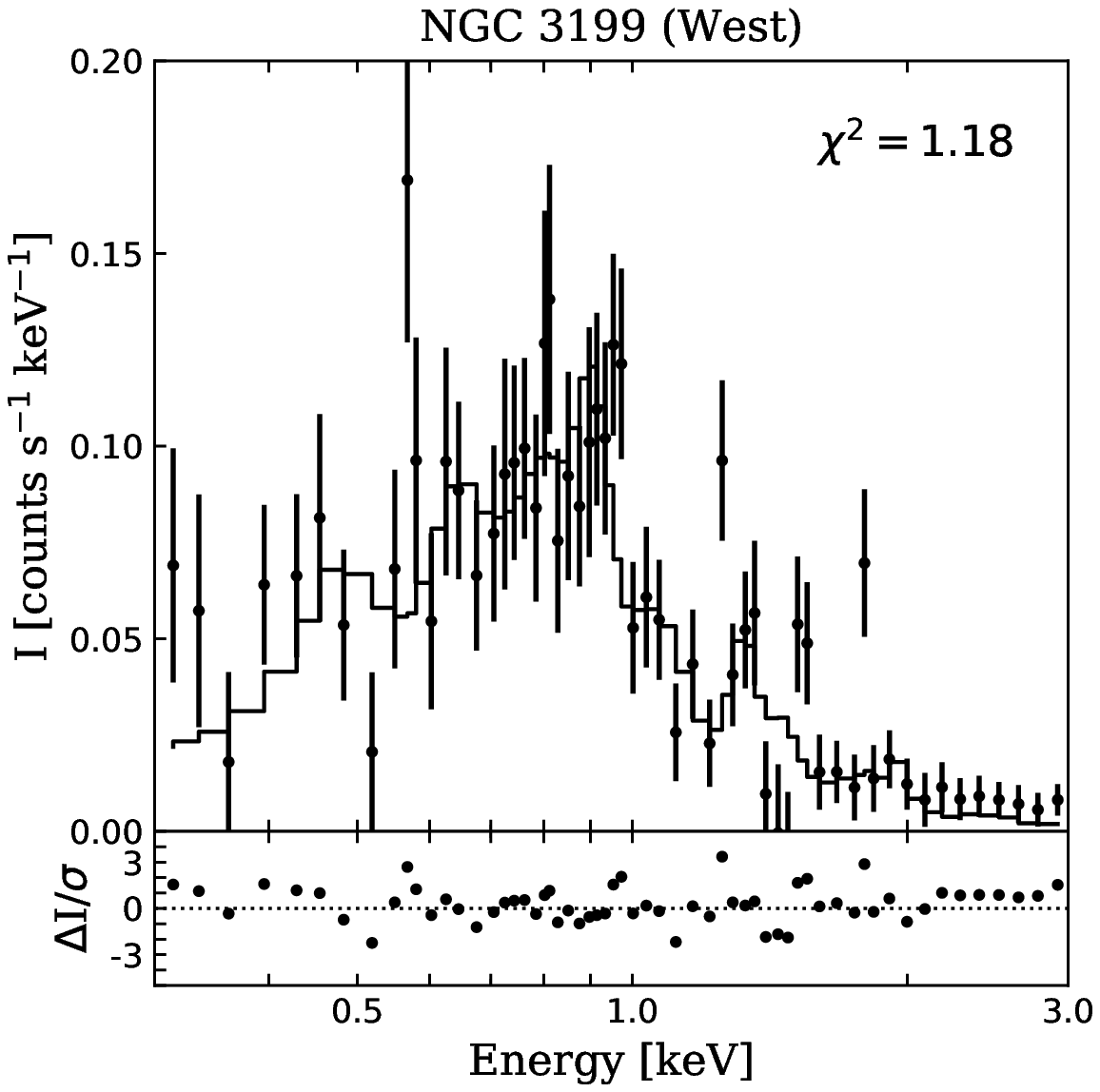}
\label{fig:ngc3199_spectra}
\caption{Background-subtracted spectra of the X-ray emission from
  NGC\,3199. Top-left panel: Diffuse X-ray emission within
  NGC\,3199. Top-right: EPIC spectra of WR\,18. Bottom panels present
  the EPIC-pn spectra of the west and east regions as defined in
  Figure~2. The diffuse X-ray spectra (NGC\,3199, East, and West) have
  been binned to 200~counts per bin for presentation. The best-fits to
  the data are shown with solid lines.}
\end{center}
\end{figure*}

\begin{deluxetable*}{cccccc}
\tablecaption{Abundance estimates of NGC\,3199}
\tablehead{
\multicolumn{1}{c}{Element} &
\multicolumn{1}{c}{$X_{\odot}$}  &
\multicolumn{1}{c}{$X/X_{\odot}$ - WN star}&
\multicolumn{1}{c}{$X/X_{\odot}$ - NGC\,3199}&
\multicolumn{1}{c}{East}&
\multicolumn{1}{c}{West}\\
\multicolumn{1}{c}{} &
\multicolumn{1}{c}{\citep{AG1989}}  &
\multicolumn{1}{c}{\citep{vdHucht1986}}&
\multicolumn{1}{c}{\citep{Stock2011}}&
\multicolumn{1}{c}{}&
\multicolumn{1}{c}{}
}
\startdata
 He & 9.77$\times$10$^{-2}$ & 9.52 & 0.91 & 0.91                 & 0.91   \\
 C  & 3.63$\times$10$^{-4}$ & 0.33 & ---  & 1.00                 & 1.00   \\
 N  & 1.12$\times$10$^{-4}$ & 52   & 0.52 & 0.52                 & 5.00$^{+10.30}_{-3.20}$\\ 
 O  & 8.51$\times$10$^{-4}$ & 0.32 & 0.53 & 0.53                 & 0.53   \\
 Ne & 1.23$\times$10$^{-4}$ & 4.96 & 1.38 & 1.38                 & 1.38   \\
 Mg & 3.80$\times$10$^{-5}$ & 5.36 & ---  & 4.01$^{+0.87}_{-0.78}$ & 4.37$^{+2.67}_{-2.46}$\\
 Si & 3.55$\times$10$^{-5}$ & 5.66 & ---  & 3.20$^{+0.83}_{-0.73}$ & 3.40$^{+5.60}_{-2.73}$\\
 S  & 1.62$\times$10$^{-5}$ & 2.93 & 1.20 & 1.20                 & 1.20  \\
\enddata
\label{tab:abundances}
\end{deluxetable*}

To study the physical properties of the X-ray-emitting gas in
NGC\,3199, we have extracted different spectra: i) a spectrum that
includes the extension of the main cavity as seen by the [O\,{\sc
    iii}] narrow-band emission, in order to estimate global properties
of the diffuse X-ray emission, ii) spectra from smaller apertures to
study variations in physical properties of the hot gas in the nebula,
and iii) a spectrum of the central star WR\,18. Three different
polygonal apertures have been defined as shown in Fig.~2. We label the
different spectra as NGC\,3199, West, East, and WR\,18.

All identified point-like sources have been removed prior to the
spectral extraction. We only used the EPIC-pn spectra of the diffuse
X-ray emission given the superior quality as compared to the EPIC-MOS
spectra. In the case of the central star, however, the smal numbers of
counts forced us to use the three EPIC spectra for a better
determination of the physical parameters (see below). Figure~4
presents all background-subtracted spectra.

WR nebulae are mainly located in the Galactic Plane, where the large
absorption column density and significant background emission
\citep[e.g.,][]{Snowden1997} pose significant difficulties for the
study of their extended X-ray emission. The spectrum of the background
extracted from the region near the camera edges outside of NGC\,3199,
as defined in Fig.~2-left, has a significant contribution in the
0.3--3.0~keV energy range (Figure~5), where the diffuse X-ray emission
from NGC\,3199 mainly concentrates.  \citet{Toala2012} discussed the
possibility of using EPIC Blank Sky observations
\citep{CarterRead2007} for the case of S\,308.  By definition, these
blank fields have flat spectra, but they do not adequately model the
local Galactic background. While useful for the extraction of spectra
of extragalactic objects, the Blank Sky observations can not be used
for objects placed in the Galactic Plane \citep[see also figure~3 and
  section 4.1 in][]{Toala2014}. Therefore, the background spectrum
extracted from the region surrounding NGC\,3199 has been used.

\section{Results}

All spectra have been modelled using XSPEC v.12.9.0 \citep{Arnaud1996}
with a two-temperature {\it vapec} plasma emission model using a {\it
  tbabs} absorption model \citep[][]{Wilms2000}. The resultant model
spectra were compared with the observed spectra in the 0.3--3.0~keV
energy range. We requested a minimum of 50 counts per bin for
the spectral fit.

\begin{figure}
\begin{center}
\includegraphics[angle=0,width=1\linewidth]{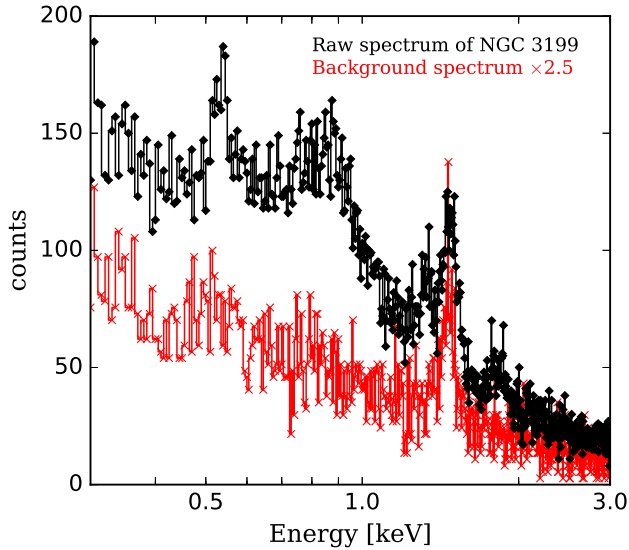}
\label{fig:ngc3199_background}
\caption{Background-unsubtracted raw EPIC-pn spectrum of NGC\,3199
  (black) and scaled EPIC-pn background spectrum (red). The prominent
  line at 1.5~keV in both spectra is the Al-K instrumental line.}
\end{center}
\end{figure}

\subsection{Global properties of the diffuse X-ray emission in NGC\,3199}

Figure~4-top left panel shows the background-subtracted EPIC-pn
spectrum of the extended X-ray emission in NGC\,3199. The spectrum is
soft with emission mainly below 3.0~keV. The peak emission comes from
energies around 0.7--1.0~keV which can be due to the Fe-complex and/or
Ne\,{\sc ix} lines. A secondary narrow peak is detected for energies
below 0.6~keV which may correspond to the 0.58~keV O\,{\sc vii}
triplet. Other lines such as those coming from the He-like Mg\,{\sc
  xi} at 1.36~keV and the Si\,{\sc xiii} at 1.86~keV can also be
identified. The diffuse X-ray emission cout rate is
340$\pm$10~counts~ks$^{-1}$ ($\approx$12,200$\pm$300~counts) in the
0.3--3.0~keV energy range.

Following the analysis carried out for the X-ray emission in other WR
nebulae, we modelled the diffuse X-ray emission in NGC\,3199 using
nebular abundances. As a first attempt we used the abundances reported
by \citet[][see Table~1]{Stock2011} and a fixed absorption column
density of $N_\mathrm{H}$=5.35$\times$10$^{21}$~cm$^{-2}$
\citep[consistent with the averaged optical extinction
  $A_\mathrm{v}$=2.92;][]{vdHucht2001}. Other elements not reported by
\citet{Stock2011}, such as C, Mg, and Si, were initially fixed to
their solar values \citep{AG1989}.

The first model resulted in a statistically good fit ($\chi^{2}$=1.24)
with temperatures of $kT_{1}$=0.18~keV and $kT_{2}$=4~keV, but the
model was not able to fit the emission lines adequately. Better fits
to the data were achieved by letting the Mg and Si abundances vary as
free parameters. The best-fit model ($\chi^{2}$=1.18) has plasma
temperatures of $kT_{1}$=0.10$^{+0.01}_{-0.03}$~keV ($T_{1} \approx
1.2 \times 10^{6}$~K) and $kT_{2}$=0.72$^{+0.04}_{-0.03}$~keV ($T_{2}
\approx 8.5 \times 10^{6}$~K) and abundances of Mg and Si 3.8 and 3.0
times their solar values. The normalization parameters\footnote{The
  normalization parameters is definedd in XSPEC as $A=10^{-14}
  \int{\frac{n_\mathrm{e} n_\mathrm{H} dV}{4 \pi d^{2}}}$, where
  $n_\mathrm{e}$, $n_\mathrm{H}$, $d$, and $V$ are the electron and
  hydrogen densities, distance, and volume, respectively.} of each
component are $A_{1}=7.50\times10^{-2}$~cm$^{-5}$ and
$A_{2}=8.95\times10^{-4}$~cm$^{-5}$. The absorbed and intrinsic fluxes
are
$f_\mathrm{X}$=(1.10$\pm0.10$)$\times$10$^{-12}$~erg~s$^{-1}$~cm$^{-2}$
and
$F_\mathrm{X}$=(4.40$\pm0.40$)$\times$10$^{-11}$~erg~s$^{-1}$~cm$^{-2}$. We
note that the contribution of the second plasma component to the total
unabsorbed flux is $\sim$8\%.

At a distance of 2.2~kpc and taking an averaged radius of 10\arcmin,
the estimated X-ray luminosity and electron density are
$L_\mathrm{X}$=2.6$\times$10$^{34}$~erg~s$^{-1}$ and
$n_\mathrm{e}$=0.3~cm$^{-3}$.

We also tried other fits with O abundance as a free parameter taking
into account the evident presence of the 0.58~keV O\,{\sc vii} triplet
in the spectrum. Even though this line appears strong, the O abundance
converged to a value close to that reported for nebular abundances by
\citet{Stock2011} , as listed in Table~1. Other fits allowing N and Ne
to vary also converge to their nebular values. Thus, the N, O, and Ne
were fixed to these values. We also tried models allowing the
absorption column density to vary, but no significant differences were
found. Furthermore, we note that another model was attempted using
abundances of a WN star \citep[][]{vdHucht1986}, as listed in Table~1,
but did not produce a good fit ($\chi^{2}>2$).

In order to assess if the presence of the Mg and Si emission lines is
real, and not an artifact of an inadequate background subtraction
\citep[see section~4.1 and figure~5 in ][]{Toala2012}, we show in
Figure~5 the background-included spectrum of NGC\,3199 and the
background spectrum extracted from the EPIC-pn data for regions
defined in Fig.~2. Whilst both spectra clearly show the instrumental
Al-K emission line at 1.5~keV, only the spectrum from NGC\,3199 shows
the excess at $\sim$1.4~keV and $\sim$1.8~keV corresponding to the
Mg\,{\sc xi} and Si~{\sc xiii} emission lines, respectively. Thus, we
are confident that these emission lines correspond to Mg- and
Si-enriched material.

Finally, we tried to evaluate the contribution of the background X-ray
emission (unresolved stars, background galaxies, ...)  projected onto
NGC\,3199 by including an additional thermal component with
temperature of 1~keV \citep[see ][]{Townsley2011}. The results of this
spectral fit indicate that this component has a small contribution but
does not change dramatically the parameters of the best-fit
two-temperature model described above. Its main effect is a reduction
of flux of the second plasma component from $\sim$8\% to $\sim$5\% of
the total flux.

\subsection{Variations in the physical properties of the
  X-ray emission in NGC\,3199}

In order to study the existence of variations in the plasma properties
in NGC\,3199, we extracted spectra from two different regions. The
background-subtracted spectra of the eastern and western regions as
defined in Figure~2 are shown in the bottom panels of Figure~4. The
count rates for the eastern and western regions are
260$\pm$10~counts~ks$^{-1}$ and 90$\pm$5~counts~ks$^{-1}$,
corresponding to detections of 9,100$\pm$300~counts and
3,200$\pm$160~counts, respectively. The spectrum extracted from the
eastern region shows very similar features as that obtained for the
global spectrum. This is consistent with the fact that most of the
diffuse X-ray emission comes from this region and the global spectrum
will be dominated by its properties. On the other hand, the spectrum
of the western region only shows a hint of the emission lines beyond
1~keV.

Before proceeding to the spectral analysis of the eastern and western
regions, we used the SciPy PYTHON Kolmogorov-Smirnof (KS) statistics
tests to evaluate similarities between spectra. This routine returns
two values, the KS statistics and its significance ($p$-value). These
parameters test the null hypothesis that the two samples were drawn
from the same distribution: small (close to zero) KS statistics or
large $p$-value means that the null hypothesis can not be
rejected. The KS statistics between the west and east spectra is 0.47
with a significance of 6.76$\times$10$^{-24}$. Therefore, we can
easily reject the null hypothesis expecting that the physical
properties differ between the two regions.

The NGC\,3199 East and West spectra were initially modelled using the
nebular abundances reported by \citet{Stock2011} with a fixed
absorption column density of
$N_\mathrm{H}$=5.35$\times$10$^{21}$~cm$^{-2}$, but we set the
abundances of N, Ne, Mg, and Si as free parameters to improve the
fits.

The best-fit model of the eastern region resulted in a reduced
$\chi^{2}$=1.23 with best-fit parameters consistent with those found
for the global spectral fit. The main plasma components are
$kT_\mathrm{1,EAST}$=0.11$^{+0.01}_{-0.02}$~keV ($T_\mathrm{1,EAST}
\approx 1.3 \times 10^{6}$~K) and
$kT_\mathrm{2,EAST}$=0.71$^{+0.04}_{-0.09}$~keV ($T_\mathrm{2,EAST}
\approx 8.2 \times 10^{6}$~K) with Mg and Si abundances of 4.0 and 3.2
times their solar values whilst N and Ne converged to their nebular
values. The unabsorbed flux in the 0.3--3.0~keV energy range is
$F_\mathrm{X,EAST}$=(2.70$\pm0.40$)$\times$10$^{-11}$~erg~s$^{-1}$~cm$^{-2}$.

In the case of the western region, the best-fit model resulted in a
reduced $\chi^{2}$=1.18 with a main plasma temperature of
$kT_\mathrm{1,WEST}$=0.20$^{+0.04}_{-0.02}$~keV ($T_\mathrm{1,WEST}
\approx 2.3 \times 10^{6}$~K) and N, Mg, and Si abundances of 5.0,
4.4, and 3.4 times their solar values. Unfortunatelly, XSPEC had
trouble fitting the second plasma component and pointed out at the
presence of a very hot plasma temperature ($kT_\mathrm{2,WEST}>2$~keV)
that contributes to less than 6\% of the unabsorbed flux of the model.
The unabsorbed flux is
$F_\mathrm{X,WEST}$=(5.90$\pm1.50$)$\times$10$^{-12}$~erg~s$^{-1}$~cm$^{-2}$.

It is interesting to note the differences between the two regions: the
western region has an enhanced N abundance in a similar way as that
defined for WN stars (see Column~1 in Table~1) with higher plasma
temperature, whilst the eastern region has N abundance closer to that
found for the optical nebula with lower plasma
temperature. Nevertheless, both regions present relatively similar Mg
and Si abundances, 3--4 times their solar values, being larger for the
western region.

\subsection{X-rays from WR\,18}

The central star of NGC\,3199, WR\,18, is detected by all three EPIC
cameras. Figure~4-top right panel shows the EPIC-pn, EPIC-MOS\,1, and
EPIC-MOS\,2 background-subtracted spectra of this WR star. X-ray
emission from WR\,18 is mainly detected in the 0.5--4.0~keV energy
range with strong features around $\sim$1.5~keV and $\lesssim$2~keV
\citep[see also figure~5 in][]{Skinner2010}. The final count rates
from the pn, MOS1, and MOS2 cameras in the 0.4--4.0~KeV range is 4.28,
1.78, and 2.0~counts~ks$^{-1}$, respectively.

To give a good characterization of the physical parameters from
WR\,18, we fitted simultaneously the three EPIC spectra using a
two-temperature {\it vapec} model as for the diffuse X-ray
emission. We followed the analysis by \citet{Skinner2010} for the {\it
  Chandra} observations of WR\,18 and used abundances as those defined
for a WN-type star \citep[see column~3 in
  Table~1;][]{vdHucht1986}. The best-fit model ($\chi^{2}$=0.87) has
an absorbing hydrogen column density of
$N_\mathrm{H}=(1.27^{+0.80}_{-0.30})\times$10$^{22}$~cm$^{-2}$ and
plasma components with temperatures of
$kT_{1}$=0.52$^{+0.23}_{-0.28}$~keV and
$kT_{2}$=1.9$^{+0.8}_{-0.8}$~keV, respectively\footnote{Note that the
  extinction toward WR\,18 is higher than that estimated for the
  diffuse X-ray emission.  This is a known issue and it is accepted to
  be due to self-absorption in the stellar wind \citep[see, e.g.,][for
    the case of WR\,6]{Oskinova2012}.}. The corresponding absorbed
(unabsorbed) flux is
3.5\,(28.2)$\times$10$^{-14}$~erg~cm$^{-2}$~s$^{-1}$, which
corresponds to a luminosity of 1.6$\times$10$^{32}$~erg~s$^{-1}$.

We also investigated the light curves of WR\,18 obtained from the
three EPIC cameras. We did not find any variation in the flux from
WR\,18 in the 0.5--4.0~keV energy range with the current {\it
  XMM-Newton} observations. We also produced soft (0.5--1.2~keV) and
hard (1.5--4.0~keV) light curves and found similar results.

\section{DISCUSSION}

The deep optical images of NGC\,3199 presented here unveil in
unprecedented detail its true extension. Although the optical nebula
has stronger H$\alpha$ emission towards the west of WR\,18 (the
southwest arc), it is in a shell that almost completely surrounds the
star. Figure~1 shows that the nebula has an approximate diameter of
$\gtrsim$20\arcmin , with WR\,18 displaced $\sim$4.7\arcmin\, from the
center toward the west. This is in sharp contrast to the 4.8\arcmin\,
radius suggested by \citet{Stock2010} from the inspection of H$\alpha$
narrow-band images from the Southern H$\alpha$ Survey
\citep{Drew2005,Parker2005}. At a distance of 2.2~kpc, the physical
size of NGC\,3199 is $\gtrsim$7~pc in radius. Another new
morphological feature is the presence of a radial fan of emission
protruding from the main nebula toward the west, and the radial rays
of emission can be traced backward to WR\,18 (see Fig.~1 upper right
panel). This feature most likely results from shadowing instabilities
\citep[e.g.,][]{Williams1999,Arthur2006}: the dense western arc
fragments into dense clumps and the UV flux from WR\,18 passes through
gaps between clumps to produce the radial features.

To the south-east we find more extended X-ray emission which appears
bounded by [O\,{\sc iii}] thin curved structures as in the other WR
nebulae studied in X-rays \citep[S\,308, NGC\,2359, and
  NGC\,6888;][]{Toala2012,Toala2015,Toala2016}. Extended [O\,{\sc
    iii}] structures have been seen in a number of WR bubbles
\citep[][]{Marston1995, Gruendl2000}. The accepted scenario of this is
that the wind from the central star and the pressurised hot bubble are
pushing through surrounding denser media, such as an ejecta shell from
an earlier evolutionary stage (RSG or LBV) or the ISM. The hot gas
fills preferencially regions of more rarified gas densities and
[O\,{\sc iii}] emission arises along the edges.

The strongest optically-emitting part of NGC\,3199, the southwest arc,
is actually seen to be in a direction from the WR star that is almost
completely orthogonal to the direction of the proper motion of the
star. As noted by \citet{Marston2001}, it is therefore not possible to
reconcile the nebular morphology with the bow shock scenario of
formation proposed by \citet{Dyson1989} in which the star has a proper
motion toward the south-west. Our deeper images clearly show a more
enclosing structure than that expected for a bow shock, with WR\,18
off-center within this extended bubble structure.

Around 25\% of massive stars are believed to have been ejected from
their parent clusters and are {\it runaway} stars
\citep[][]{Blaauw1961,Fujii2011,Gies1987,Hoogerwerf2000}. Indeed, a
runaway scenario could well explain any bow shock forming in the
direction of the motion of the star. In order to check whether WR\,18
is a runaway star, we need to determine if it has a motion that is
distinct from its surroundings and possibly in the direction of the
putative bow shock, the south-western arc of NGC\,3199.

The proper motion of WR\,18 noted in \citet{Marston2001} was derived
from {\it Hipparcos} measurements. More recently, {\it Gaia}
measurements in the DR1 release have become available
\citep[][]{Gaia2016}. The Tycho-Gaia Astrometric Solution dataset
(TGAS) of the {\it Gaia} DR1 release provides improved proper
motions. The results continue to be consistent with the {\it
  Hipparcos} results (a proper motion toward the north-west direction;
see Fig.~1) but with much smaller error. The overall proper motion is
($\mu_{\alpha}$,
$\mu_{\delta}$)$_\mathrm{Gaia}$=($-$5.98$\pm$0.17~mas~yr$^{-1}$,
3.48$\pm$0.16~mas~yr$^{-1}$).

Within a radius of 20\arcmin\, around WR\,18 there are 93 TGAS
stars. Typical errors quoted for the proper motion of these stars are
1.5~mas~yr$^{-1}$ in R.A. and 1.2~mas~yr$^{-1}$ in Dec. If we consider
all stars within 2 times this error value from the proper motion of
WR\,18 we find a total of 44 TGAS stars, or half of all TGAS stars in
the region. In other words, there is a significant grouping of stars
in the part of the sky containing WR\,18 that have similar proper
motions, both in terms of size and direction. To help illustrate this,
we show in Fig.~3 the position of the TAGS stars that lie in the
vicinity of our nebular images in white circles.

We may conclude that the proper motion of WR\,18 gives no indication it is
moving at some abnormal, runaway velocity with respect to its
surroundings. Indeed, the cluster Westerlund~2 is found 57\arcmin\,
away to the east and has a proper motion of ($\mu_{\alpha}$,
$\mu_{\delta}$)$_\mathrm{Gaia}$=($-$6.77~mas~yr$^{-1}$,
4.84~mas~yr$^{-1}$) with a total error of $0.24$~mas~yr$^{-1}$
\citep{Kharchenko2013}. Thus, the simplest interpretation of the
proper motion of WR\,18 is that it is part of the bulk Galactic motion
of stars in the spiral arm of the Galaxy that contains both WR\,18 and
Westerlund~2.

{\it Herschel} PACS images \citep[][]{Pilbratt2010,Poglitsch2010} of
the region also show that there is a large shell around WR\,18 with
WR\,18 towards the western edge (see Fig.~6) but not toward the
direction of the proper motion reported by {\it Hipparcos} or {\it
  Gaia}. The overdensity of materials to the west of the star would
then naturally be caused by a pile-up of materials as the wind from
WR\,18 sweeps material up against the western edge of this large
shell. The NGC\,3199 nebula exhibits a particularly warm region of
dust in the {\it Herschel} 100/160~$\mu$m ratio image shown in
Figure~6. This is most likely due to the dust in this part of the
nebula that is heated by the stron UV flux from WR\,18. Although the
[O\,{\sc iii}] 88 micron emission line is on the edge of the PACS
broadband 100~$\mu$m spectral bandpass, \citet{Whitehead1988}
concluded that the main emission mechanism in this region is due to
radiative ionization from the central star and not by shocks (as it
would be the case of a bow shock scenario).

To ensure that WR\,18 is the main source of ionization in NGC\,3199,
we searched for other OB stars in the vicinity of WR\,18. We took the
44 TGAS stars with similar proper motion as WR\,18 and seached for
their spectral type. None of these stars have an O spectral type. We
found only one BOV star (CD$-$57$^{\circ}$3120) located 4.6\arcmin\,
from WR\,18 (black circle in Fig.~3) projected on the bright H$\alpha$
arc. Nevertheless, WR\,18 will dominate the physics of NGC\,3199. Its
flux is almost 200 times larger than that estimated for
CD$-$57$^{\circ}$3120 \citep[assuming standard stellar parameters for
  a B0V star from][]{Cox2000}.

Our inescapable conclusion is that WR\,18 is not a runaway star and is
likely to have been formed not far from its current position. The
differences in spectral properties of the X-ray-emitting plasma and
the reported variations in abundances are due to a combination of the
initial inhomogeneous configuration of the ISM, which triggered
different instabilities stirring the material unevenly, and the
current metal-enriched fast stellar wind from WR\,18.

\begin{figure}
\begin{center}
\includegraphics[angle=0,width=1\linewidth]{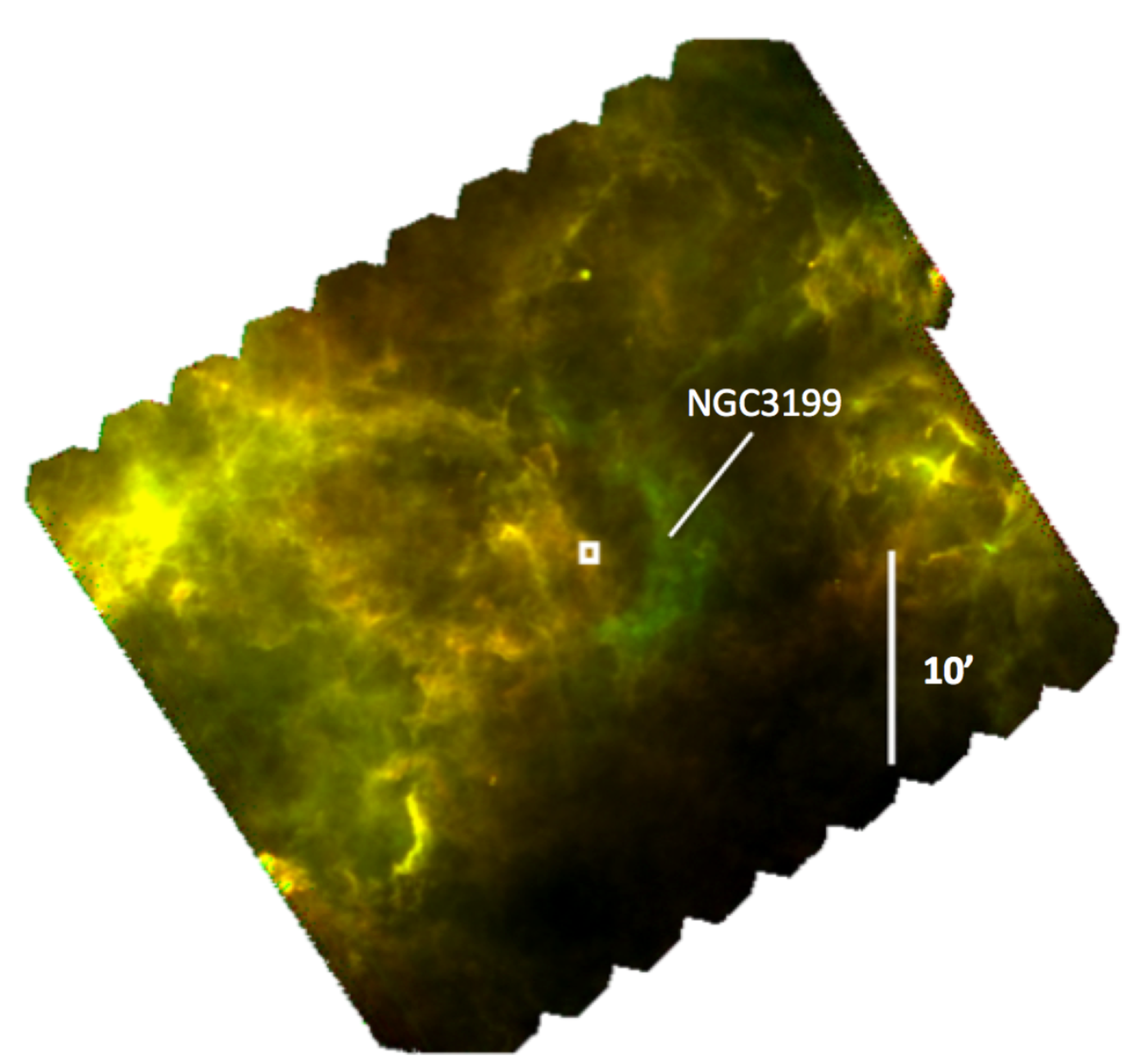}
\label{fig:WR18_Herschel}
\caption{Two color {\it Herschel} PACS (Ob.\,ID 1342249269; PI:
  A.\,Marston) far infrared image at 100~$\mu$m (green) and 160~$\mu$m
  (red) wavelengths. The main nebula NGC\,3199 exhibits an arc of high
  100 to 160$\mu$m fluxes indicative of hot dust. The position of WR18
  is shown by the white box at centre. North is up and east is to the
  left.}
\end{center}
\end{figure}

Optical emission-line studies of the chemistry of NGC\,3199 have shown
some apparent contradictory information between them. There
appear to be some N-enhanced ejecta-type of materials but there have
also been reports of ISM abundances in NGC\,3199. In our scenario it
would be clear that most materials in the main part of the nebula
should be associated with ISM materials that are from the giant shell
inside of which exists WR\,18. But WR stars can typically have a
clumpy ejecta phase, e.g. RSG or LBV phase. Ejected materials during a
RSG or LBV phase is N-enriched \citep[e.g.,][]{Stock2014}. Clear
evidence of metal-enriched materials is also seen close to WR\,18
\citep{Marston2001} and up to several parsecs away – which is a
typical size for WR ring nebulae following an ejecta phase
\citep[e.g.,][]{Toala2011}.

WR stars exhibit highly-ionized species in their X-ray spectra, in
particular those classified as WN stars \citep[e.g.,
][]{Skinner2010,Skinner2012}. Recently, \citet{Hue2015} presented the
most detailed study of the X-ray spectrum from a WN4 type star,
WR\,6. These authors used deep {\it Chandra}/HETGS observations to
obtain a very high-resolution X-ray spectrum and reported the presence
of a number of H-like and He-like emission lines, including strong
lines of Mg\,{\sc xi} at 1.36~keV and Si\,{\sc xiii} at
1.86~keV. \citet{Hue2015} concluded that the current wind of WR\,6 is
ejected in a constant spherical expansion and that X-rays emerge from
regions within 30--1000 stellar radii. If this would be also the case
for WR\,18, it would help explain the high abundances of N, Mg, and Si
in NGC\,3199, specifically in the region around the star (e.g., the
eastern region) and would imply that mixing with the ISM has been less
efficient in that region.

Finally, it is interesting to speculate that the passage of a slowly
expanding giant shell triggered the formation of the progenitor O star
of WR\,18. An expansion rate of a few km~s$^{-1}$ is all that would be
needed for the shell to advance to the current position of NGC\,3199
in the million years or so that WR\,18 has taken to evolve to a WR
star. WR\,18 would then be the result of a large clump that was
triggered into star formation by a passing wave of material. There are
numerous examples of triggered clumps of this kind in {\it Herschel}
observations \citep{Hill2011,Rivera2015,Zavagno2010}.

\section{Summary and conclusions}

The deep optical narrow-band images (H$\alpha$, [O\,{\sc iii}], and
[S\,{\sc ii}]) presented here, unveiled the true extension of the
Wolf-Rayet nebula NGC\,3199. The WR nebula around WR\,18 has an
elongated shape with 18\arcmin$\times$22\arcmin\, in size with its
central star off-centered 4.7\arcmin\,toward the west. The analysis of
the narrow-band images of NGC\,3199 show a complex structure of
radially-distributed filaments pointing outwards from WR\,18.

We presented the {\it XMM-Newton} discovery of the diffuse X-ray
emission toward NGC\,3199. These observations render NGC\,3199 the
fourth WR nebula detected in X-rays. The current observations show
that the diffuse X-ray emission is delimited by the [O\,{\sc iii}]
narrow-band emission (as in the cases of S\,308, NGC\,2359, and
NGC\,6888 around WR6, WR7, and WR136) with a maximum toward the
south-east region.

The global X-ray properties of NGC\,3199 are similar to those found in
other WR nebulae. The estimated intrinsic X-ray flux in the
0.3--3.0~keV energy band is
$F_\mathrm{X}$=(4.40$\pm0.40$)$\times$10$^{-11}$~erg~s$^{-1}$~cm$^{-2}$,
which corresponds to an luminosity of
$L_\mathrm{X}$=2.6$\times$10$^{34}$~erg~s$^{-1}$ at a distance of
2.2~kpc. The dominant plasma temperature is
$T\approx$1.2$\times$10$^{6}$~K with a hotter component that
contributes less than 8~per cent of the unabsorbed flux. The estimated
electron density is $n_\mathrm{e}$=0.3~cm$^{-3}$.

A careful analysis of the X-ray properties revealed temperature and
abundances variations within the nebula: regions close to the main arc
(the western region) are dominated by hotter gas with enhanced N, Mg,
and Si abundances pointing to the current role of WR\,18 in heating
and enriching NGC\,3199. The eastern region has lower plasma
temperature with abundances similar to those reported previously for
the nebular material implying that mixing is more important in this
region. We suggest that these high abundances are due to the current
metal-rich wind from WR\,18 as dectected in other WN4 stars (e.g.,
WR\,6).

With the help of the {\it Gaia} first release and {\it Herschel}
images we conclude that WR\,18 is not a runaway star and it is more
likely that the current shape of NGC\,3199 is due to the initial
inhomogeneous configuration of the ISM.

Finally, the properties derived from our {\it XMM-Newton} EPIC
observations of WR\,18 are in agreement to those reported previously
based on {\it Chandra} observations. We further analyzed the X-ray
light curves of WR\,18 as obtained from the three EPIC cameras and we
found no evidence of variations over timescales of the current
observations ($<50$~ks).

\section*{Acknowledgements}

We would like to thank the anonymous referee for helpful suggestions
that improved the presentation of our results. This work was based on
observations obtained with {\it XMM–Newton}, {\it Herschel}, and {\it
  Gaia} satellites. {\it XMM-Newton} is an ESA science missions with
instruments and contribution directly funded by ESA Member States and
NASA. {\it Herschel} is an ESA space observatory with science
instruments provided by European-led Principal Investigator consortia
and with important participation from NASA. The {\it Gaia} data have
been processed by the {\it Gaia} Data Processing and Analysis
Consortium (DPAC). Funding for the DPAC has been provided by national
institutions, in particular the institutions participating in the {\it
  Gaia} Multilateral Agreement.

The authors thank Don Goldman (don@astrodon.com) for providing the
narrow-band [S\,{\sc ii}], H$\alpha$ and [O\,{\sc iii}] images of
NGC\,3199. We thank G.\,Ramos-Larios for helping process the optical
images. MAG acknowledges support from the grant AYA 2014-57280-P,
co-funded with FEDER funds.


\end{document}